\documentclass[twoside,fleqn]{article}
\usepackage{espcrc2,epsf}

\newcommand{\AmS}{{\protect\the\textfont2
  A\kern-.1667em\lower.5ex\hbox{M}\kern-.125emS}}

\newif\ifcolor
\colortrue

\title{Cost-Effective Clustering
}

\author{
        Steven Gottlieb\\
{Department of Physics, Indiana University, Bloomington, IN 47405, USA}
} 
       
\begin{document}

\begin{abstract}
Small Beowulf clusters can effectively serve as personal or
group supercomputers.  In such an environment, a cluster can be optimally
designed for a specific problem (or a small set of codes).
We discuss how theoretical analysis of the code and benchmarking on 
similar hardware lead to optimal systems.

PACS/Keywords: 12.38.Gc, Lattice QCD, Linux, Beowulf Clusters, Supercomputing
\end{abstract}

\maketitle

\section{INTRODUCTION}
The Beowulf project began in 1994 at the NASA Goddard Space Flight Center.
The history, current status and a list of about 100 clusters may
be found at http://www.beowulf.org.
This year at Supercompting 2000, Gordon Bell sponsored a prize for
a computer constructed or purchased for under \$10,000.
Thus, cost-effective clustering is a timely topic.

I assume you need to do a lot of computation and need to get it done 
inexpensively rather than in the shortest time.  It helps
if you can run more than one job at a time.  
This should be contrasted with 
weather prediction, where being
able to run seven one-day predictions that each take seven days is not
very useful.  For weather we must exceed a specific minimal speed that will
let us predict tomorrow's weather from current conditions.
The cost model presented here
is based on capital expenditure for hardware.  A more sophisticated
model would include maintenance, electricity, cooling, floor space, etc.

These design principles are advocated: 

1) Know the bottlenecks and requirements of your problem.
With this knowledge, you can avoid building an unbalanced system that,
for instance, might put too much money into a network whose high performance
is not required.
2) Design for the sweet spot. (Note: the sweet spot changes with time and may
depend on the problem.)  As an example, the highest density memory available
at any time tends to be quite expensive.  Currently, it is less expensive to 
buy two 256 MB parts than one 512 MB part.
3) Design for total system cost effectiveness.  If a 10\% increase in the
speed of the processor results in a 5\% performance increase, since
the processor is only one component, the system price
increase might be less than 5\%.
4) Benchmark as much as you can before deciding on a design.

A cluster compute node can be very simple.  Six items
are (almost) mandatory: motherboard, CPU, memory, network card, case and
floppy drive.  The first three items will be the most expensive if using
a FastEthernet network (unless you insist on using rack mounted cases).  
Otherwise, the network card will be a major expense.
Some motherboards come equipped with a FastEthernet interface.  I have
also seen some systems without individual cases.  A hard drive or video
card may be useful.  Some Beowulf designers discourage a hard drive on
each compute node.  If your application requires lots of scratch disk i/o,
then you may need a disk on the node.  When a node repeatedly fails to
reboot, a video card can help diagnose the problem.  (At IU in two years of
running with 40 nodes, this has only been necessary about half a dozen times.)
We have a few spare video cards to install when needed.

The Indiana University Physics Department received \$50,000 in 1998
to build a 32-node Linux cluster.  The machine we built in Nov. 1998
is called CANDYCANE, which stands for CPUs And Network Do Your Calculation And
Nothing Else.  CANDYCANE is an appropriate name because it was designed for
the ``sweet spot,'' that is, components were picked to give the best 
price-performance ratio attainable.  
The cost per node was \$693 for a Pentium II 350, with a 4.3
GB hard drive and 64 MB of ECC RAM.  Each node has a floppy drive and a
FastEthernet card.  
The 40-port HP Procurve switch cost
about \$2,000, so the total cost was about \$25,000.
In November, 2000
it would have been possible to build this system for $<$\$320 per node, or
for approximately \$12,000.  An
even more attractive alternative would be a diskless Athlon 600 MHz
system for which the per node cost is about \$275.  This node would have
much better performance than the PII 350; however, the FastEthernet would
be a bottleneck on the MILC code with Kogut-Susskind quarks.  Even so,
a 32 node system with a minimum performance of 1280 and 1660 Mflops, for $8^4$
and $14^4$ sites per node, respectively, could be built for under \$10,500.
This works out to a cost/MF of between \$6.3 and \$8.2.

In Sec.~2, we describe the key issues for good performance.  Section 3
gives details of single node performance.
Section~4 points the reader to a web site with extensive benchmarks,
gives cost estimates for several designs and compares 
cost-performance ratios for these clusters and a number of supercomputers.  
For additional information about emerging technologies for clusters see 
Ref.~\cite{APS}.

\section{KEYS TO PERFORMANCE}

A very simple approach to achieving good performance for domain
decomposition codes like Lattice QCD codes is to optimize single node
performance and to try to avoid degrading performance too much when one has to
communicate boundary values to neighboring nodes.  The key to a cost effective
design is an appropriate balance.  Floating point performance is more easily
adjustable than network performance because processors come in many speeds, but
there are only a few choices for the network.

A simple performance model of the Kogut-Susskind Conjugate Gradient
algorithm gives this bandwidth requirement to overlap communication and
floating point operations:
\begin{equation}  
MB = 48 MF/ (132 L) = 0.364 MF / L ,
\end{equation}
where $MB$ is the {\it achieved\/} bandwidth in Megabyte/s, $MF$ is the 
{\it achieved\/} floating
point speed in Megaflop/s on matrix-vector multiplication
and an $L^4$ portion of the grid is on
each node.  We assume there are neighboring nodes in each direction, {\it
i.e.}, 16 or more nodes.  The constant factor 0.364 is specific to
KS quarks.  However, the $1/L$ behavior is typical of the domain decomposition
approach to parallelism and comes from the surface to volume ratio.

\begin{figure}[t]
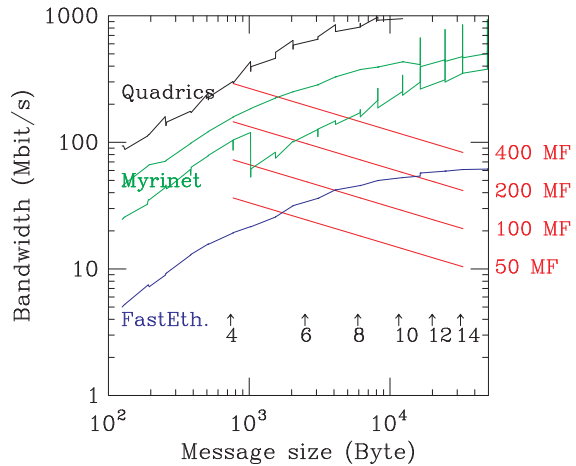

\ifcolor{\epsfxsize=0.99 \hsize\epsffile{fig1.ps}}
\else{\epsfxsize=0.99 \hsize\epsffile{fig1bw.ps}}\fi
\vspace{-28pt}
\caption{Measured bandwidth and the simple performance model.}
\label{fig:performance}
\end{figure}

Figure \ref{fig:performance} shows a log-log plot 
of measured bandwidth on a ping-pong test for
three types of hardware and the performance model for several processor speeds.
The messages vary in size from 800 bytes to 30 KB for problem sizes of
interest.  The arrows near the bottom of the graph correspond to
different L values.
The FastEthernet \ifcolor{(blue)\ }\fi and Myrinet \ifcolor{(green)\ }\fi
curves come from measured performance on the Roadrunner (RR)
supercluster at the
Albuquerque High Performance Computer Center. 
Two curves are shown for Myrinet.  With the newer drivers, bandwidth is better
and smoother.
The Quadrics curve comes from the Teracluster at Lawrence Livermore
National Laboratory (LLNL).
The measurement was done using the Netpipe program from the
Ames Scalable Computing Laboratory \cite{AMESLAB}.  
The straight \ifcolor{red\ }\fi lines come from the 
performance model presented above and
are plotted for matrix times vector speeds of 50, 100, 200 and 400 MF.
We need to run at a large enough value of L so that the measured bandwidth
is above the straight line (for whatever speed our processor achieves for
the corresponding value of L).  Because of cache effects, the processors
will achieve higher speeds when $L$ is small, but that requires the highest
bandwidth.  Thus, pushing up the communication rate for small
messages is important.  Being able to run for a small value of $L$ with
high efficiency allows running a fixed size problem at high total
performance.
We see that none of the networks achieves more than a small fraction of its peak
bandwidth for the message sizes of interest.  A system design based on
achieving that peak bandwidth would almost certainly be communication
bound, {\it i.e.}, money would have been spent on floating point capacity
that could not be used.  There are large differences in the prices of
FastEthernet, Myrinet and Quadrics hardware.  Choice of network can
obviously play a critical role in system performance and
cost-effectiveness.

\section{SINGLE NODE PERFORMANCE}

The single node performance is likely to depend upon such issues as the
quality of the CPU, the performance and size of cache(s), the bandwidth to
main memory and the quality of the compiler.  For message passing
performance, key issues are the latency,
peak bandwidth, processor overhead and the message passing software.
It is important to make the right choices when designing your system.

For the CPU, one can choose among Intel Celeron, Pentium II, PIII, PIV,
Itanium; AMD Athlon, Thunderbird, Duron; Compaq Alpha and other possibilities.
The Celeron may have limited performance because it only has a 66 MHz
front side bus (FSB).  If access to memory is important (as it is for this
application) the 100 or 133 MHz FSB of the PII and PIII will be useful.
The PIV is quite new and currently expensive.  Itanium is soon to be available
and some results are provided below.  The Alpha has great performance, but
it is expensive.

There are currently several memory types available for different processors.
They include PC100, PC133, Rambus, Double Data Rate (DDR), which is also known
as PC1600 or PC 2100.
PC100 and PC133 are fairly mature at this stage, and
there is little difference in price (currently about 0.5\$/MB).
Rambus was quite expensive at introduction, but has recently decreased 
quite a bit (currently slightly more than 1\$/MB).  
DDR is just now coming
to market (Micron), but it does not carry too much of a premium (currently
similar to Rambus).  Motherboards that use DDR memory are just coming to
market.
Pick the right amount of memory for your problems: you never want your code
to swap, but neither do you want to buy a lot of memory that you never use.

Choice of motherboard is crucial.  It must be matched to the
processor and memory.  The support chip can have an important impact 
on performance (as we explore below).
The motherboard determines the number of processors per node.  We discuss below
whether dual processor systems are more or less cost-effective for a
particular application.
The motherboard will also determine whether you have a faster or wider
PCI bus than the initial standard 33 MHz-32bit bus.  This may be important if
you have a higher speed network like Myrinet.

The last critical choice is the network hardware and software.
It properly deserves its own section, but to save space we
briefly discuss it here.
FastEthernet is the commodity network.  Other choices such a Myrinet,
Giganet, Gigabit Ethernet, Quadrics QSnet and SCI from Dolphin/Scali
have higher performance, but are quite expensive compared to FastEthernet.
(Including the card and switch, you can expect to spend about \$1,500
per node, except for Quadrics which is over \$3,000 per node.)
Because of the big jump in price and performance, it is necessary to make
sure the system maintains balance between CPU and network.
Regarding software choices,
under FastEthernet, MVIA\cite{MVIA} and GAMMA\cite{GAMMA} 
software have reduced latency compared with standard TCP/IP.
I have recently tried VMI under Myrinet which is being
developed at NCSA and found it to be
superior to running under standard GM (Myrinet supplied) driver.

\begin{figure}[t]
\epsfxsize=0.99 \hsize
\epsffile{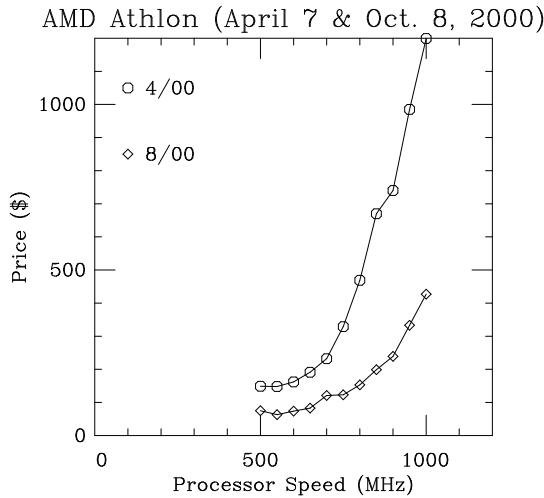}
\vspace{-28pt}
\caption{Processor price {\it vs}. speed.}
\label{fig:price_vs_speed}
\end{figure}


Turning next to single node performance, we note that it is easy to waste
a lot of money on a poor system design.  To illustrate this, we consider
the various speed AMD Athlon processors available and their prices on two
particular days.  Although we focus on Athlon here, the same considerations
apply to Intel or other processors.  Figure \ref{fig:price_vs_speed}
shows that processor price is a rapidly increasing function of speed.  
It can also be a rapidly decreasing function of time, especially for the
faster processors.


Dividing the price
by the speed of the chip, we find that the relative expense rises
rapidly for the faster chips \cite{WEBSITE}.  
On April 7, 2000, there
was an apparent sweet spot at 600 MHz.  The faster chips have a higher
price-performance
ratio.  Depending upon the costs of the other components of the system,
the entire system may have a higher or lower price-performance ratio.

For our QCD codes, access to memory is quite important.  
In our first example, we compare a Celeron chip with a 66 MHz
FSB and a Pentium II chip with a 100 MHz FSB.  
The Pentium's clock speed is only 6\% faster than the Celeron, but its
performance is 38\% faster for $L=$6 and 23\% faster for $L\ge$8.  This
is because of its larger cache and 52\% faster FSB.
When we overclock the Celeron, it has a slower FSB but a faster CPU speed
than the PII, and performance is about equal except for $L=$6 where the
smaller cache probably is the source of the difference.

By comparing 500 MHz and 600 MHz Athlons,
we demonstrate that performance does not increase in proportion to the
speed of the chip.  This is because memory access is fixed by the FSB speed.

\begin{table}[t]
\caption{Megaflop rate of Intel Processors}
\label{tab:intel}
\begin{tabular}{ccccc}
\hline
CPU& Celeron &Celeron&PII\\
Clock(MHz)& 330 &416(OC)&350\\
FSB(MHz)& 66 &82 &100 \\
\hline
L& MF&MF &MF\\
4 & 102 & 117 & 113 \\
6 & 60 & 74 & 83 \\
8 & 57 & 71 & 72 \\
10 & 57 & 71 & 70 \\
12 & 56 & 70 & 70 \\
14 & 56 & 71 & 70 \\
\end{tabular}
\caption{Megaflop rate of Athlon Processors}
\label{tab:athlon}
\begin{tabular}{ccc}
\hline
L & 500 MHz & 600 MHz \\
\hline
4 & 231 & 276 \\
6 & 129 & 135 \\
8 & 97 & 102 \\
10 & 92 & 97 \\
12 & 90 & 95 \\
14 & 89 & 93 \\
\end{tabular}
\end{table}
\begin{table*}[thb]
\caption{Megaflop rates of various motherboards and CPU combinations or cluster nodes}
\label{tab:pc133}
\begin{tabular}{ccccccc}
\hline
L & Gigabyte GA6VXE+ & Intel CC820 & SM PIIISED & SM P6SBA 
& RR \dag & LL \ddag \\
  & Pentium III 533B    & PIII 533B & PIII 533B & PII 350 
& PII 450 & PIII 733 \\
\hline
4 & 186 & 182 & 174 & 114 & 142 & 319 \\
6 & 106 & 98 & 94 & 83 & 99 & 140 \\
8 & 81 & 75 & 73 & 72 & 82 & 130 \\
10 & 76 & 72 & 70 & 70 & 79 & 127 \\
12 & 76 & 70 & 69 & 70 & 78 & 127 \\
14 & 73 & 70 & 69 & 70 & 78 & 126 \\
\hline
\multicolumn{3}{l}  \dag Roadrunner: Portland Group compiler \hfil\\
\multicolumn{3}{l} \ddag Los Lobos: Portland Group compiler \hfil\\
\end{tabular}
\end{table*}

The 600 MHz chip has a peak speed 20\% faster than the 500 MHz chip.  With
$4^4$ lattice points, we do see a 20\% speed up, but for the larger problem
sizes that do not fit into cache, there is only a 5\% speedup.  We expect
that for even faster processors, performance increases will be marginal.

Since memory access is so crucial, I have purchased a Pentium III 533B
chip that uses PC133 memory.  In theory, it should provide about 33\%
better performance than a similar chip with PC100 memory.
I have tried three different
motherboards using different support chips and the results are
disappointing.
The Gigabyte GA6VXE+ motherboard uses a VIA chip set, 
the Supermicro (SM) PIIISED
uses the Intel 810e chip set, and I also tried an Intel CC820 motherboard
using the Intel 820 chip set.  
The results are not particularly better than a PII350 or 450 MHz chip using 
a BX motherboard, except when the problem fits
in cache.  In the future, I hope to test whether the PIII prefetch instructions
can be used to improve this situation.  (This technique would also be
applicable to a BX motherboard with PIII installed.)  The next version of
the Portland Group Compiler is supposed to have better facilities for telling
the compiler when to prefetch data to cache.
(See Table~\ref{tab:pc133}.)

It is possible that the 815e support chip from Intel will provide better
performance with the 133 MHz FSB, but I have not done any tests myself.
There are benchmarks at the Intel developer's web site that look promising.
However, there are support chips from ServerWorks that support PC133
memory well.  Here are results from Los Lobos (LL), that uses 
Intel 733 MHz chips in IBM Netfinity servers that use a ServerWorks chip set. 
SuperMicro is manufacturing a dual processor
motherboard that uses a support chip from ServerWorks, but the motherboard
currently costs about \$280, which is about twice the price of
dual processor BX motherboard.  Also, this board requires registered memory
which will add to the cost of the system.

You might have noticed above that the Athlon numbers are fairly impressive
especially considering that an Athlon chip will cost less than an Intel
chip at comparable speed.  There has been good news recently on the
Athlon front.  DDR memory is now available from Micron, and there will
soon be motherboards that use the AMD 760 chipset which supports DDR
memory.  Dual processors have not been available for the Athlon, but
there will be an AMD 760MP chip that will support dual processors.
On October 10, 2000, AMD announced a demonstration dual-CPU DDR system, so
it should not be too many more months before these are available to the
consumer.

With the help of programmers at Intel, the MILC benchmarks
have been tuned and run on several Intel Itanium(tm) based systems.
The code that was run does
not include any assembler instructions, but prefetching hints to the compiler
have been inserted to enhance performance.  The Itanium(tm) processors have the
best performance we have seen for our code.

\begin{table}[bh]
\caption{Megaflop rate of Itanium Processors}
\label{tab:itanium}
\begin{tabular}{cccc}
\hline
 & 600 MHz & 667 MHz & 800 MHz\\
L & 100$\times$1 & 133$\times$2 & 133$\times$2\\
 & 2MB/1GB & 4MB/4GB & 4MB/2GB\\
\hline
4 & 692 & 761 & 916\\
6 & 646 & 726 & 867 \\
8 & 290 & 591 & 732 \\
10 & 214 & 464 & 539 \\
12 & 187 & 330 & 359 \\
14 & 178 & 301 & 326 \\
\end{tabular}
\end{table}

It should be noted that the 667MHz processor is a pre-production pilot.
The two faster processors have double-pumped front side bus
and they also have a larger cache. By looking at ratios of results
we can see the effects of cache size, external memory speed and processor
speed.  A second table and interpretation of the relative performance may
be found in the original transparencies\cite{WEBSITE}.


\section{PRICE-PERFORMANCE RATIOS \& BENCHMARKS}

The simple performance model presented above can help us predict when the
communication
and floating point are in reasonable balance, but it is no substitute for
real benchmarks.
A web site for MILC benchmarks may be found at
physics.indiana.edu/\~{}sg/milc/benchmark.html.  
All the benchmarks presented are for single precision Kogut-Susskind
conjugate gradient.  Due to limitations of space and time, we refer the
reader to the web site.

We consider bare-bones cluster nodes with either 450 MHz PII chips as in
Roadrunner, or 733 MHz PIII chips as in Los Lobos.  For the Los Lobos level
system, we are assuming a ServerWorks dual-CPU capable motherboard will have
comparable performance to the IBM Netfinity nodes.  
Each node contains 64 MB of ECC memory per processor
and a 4.3 GB hard drive.  Prices
are based on a search of www.pricewatch.com on Oct.~31, 2000.
Performance expectations are based on Roadrunner or Los Lobos.
(An AMD Athlon based system was considered in Sec.~1.)  

To build a single CPU node like Roadrunner, but with 64 MB of memory would
cost \$325.  A dual CPU node, with 128 MB would cost \$527.  If more memory
is desired, it should cost less than \$1 per Megabyte.  Los Lobos style nodes
are \$634 and \$878, for single and dual cpu, respectively.  Per port
FastEthernet switch costs are
about \$56, \$185 and \$240, for 32, 72 and 144 ports, respectively.
Myrinet cost is \$1527 per port and scales linearly up to 128 ports with
the Clos switches.  These prices are for LANai 9, Myrinet 2000 cards,
and better network performance than on Los Lobos is expected.

For a single CPU RR level system, the price-performance ratio
in \$/MF is
7.2--9.3, 10--13, $\approx$11--15 and 28--31, for FastEthernet with
32, 64, 128 nodes and Myrinet, respectively.  With dual CPUs, the
numbers are 8--11, 12--16, 14--19 and 20--23.  We see that the second
CPU makes the Myrinet based system considerably more cost effective;
however, for FastEthernet, although the marginal cost is small, the
performance gain is not that great either, and the system is less cost
effective.  With the more expensive and higher performance LL level
nodes, FastEthernet cost in \$/MF is about 11, 14 and 17, and
Myrinet is
20--22  for single CPU systems.  With a second processor the Myrinet
number drops to 16--21.  Dual CPU benchmarks have not been run with 
FastEthernet, but the network performance
should be even more of an issue here, and
we expect the cost effectiveness to be somewhat less.

\begin{table}[t]
\caption{Price-performance ratios}
\label{tab:priceperformance}
\begin{tabular}{lc}
\hline
Computer (date of quote)& \$/MF \\
\hline
RR level 1 CPU FE (10/00) & 7--15 \\
RR level 1 CPU Myrinet (10/00) & 28--31 \\
RR level 2 CPU FE (10/00) & 8--19 \\
RR level 2 CPU Myrinet (10/00) & 20--23 \\
LL level 1 CPU FE (10/00) & 11--17 \\
LL level 1 CPU Myrinet (10/00) & 20--22 \\
LL level 2 CPU Myrinet (10/00) & 16--21 \\
\hline
64-node SGI Origin 250 MHz (2/99) & 193\\
44 node Cray T3E (2/99)  & 480 \\
256 node IBM Power 3 SP (2/00) &  166 \\
 \quad with estimated discount &  91 \\
64 CPU Compaq Alpha Server SC & 150 \\
\hline
\end{tabular}
\end{table}

This work was supported by the U.S. DOE under grant DE-FG02-91ER 40661.
Special thanks to the MILC collaboration,
the Albuquerque High Performance
Computer Center, Indiana University, LLNL,
National Center for Supercomputing Applications, Pittsburgh
Supercomputer Center and San Diego Supercomputer Center.


\begin{thebibliography}{9}
\def\npproc{Nucl. Phys. (Proc. Suppl.)\ }
\def\npb{Nucl. Phys. B}
\def\prd{Phys. Rev. D}
\def\prl{Phys. Rev. Lett.\ }
\def\pl{Phys. Lett.\ }
\def\etal{{\it et al}.,\ }

\bibitem{APS}
See http://physics.indiana.edu/\~{}sg/pccluster.html\ for several
talks on cluster technology and application performance.

\bibitem{AMESLAB}
Visit http://www.scl.ameslab.gov.  Netpipe can be downloaded from
there.

\bibitem{MVIA}
To find out more about M-VIA, see http://www.nersc.gov/research/FTG/via.

\bibitem{GAMMA}
See the GAMMA web site of G.~Ciaccio, http://www.disi.unige.it/project/gamma/.

\bibitem{WEBSITE}
See http://physics.indiana.edu/\~{}sg/ccp2000/\ for this
graph and others not included here.

\end{thebibliography}
\end{document}